\documentclass[conference]{IEEEtran}
\IEEEoverridecommandlockouts
\usepackage{cite}
\usepackage{amsmath,amssymb,amsfonts}
\usepackage{algorithmic}
\usepackage{graphicx}
\usepackage{textcomp}
\usepackage{enumitem}
\usepackage{url}
\def\BibTeX{{\rm B\kern-.05em{\sc i\kern-.025em b}\kern-.08em
    T\kern-.1667em\lower.7ex\hbox{E}\kern-.125emX}}

\newcommand{\Comment}[1]{}

\begin{document}

\title{Neural Networks for Safety-Critical Applications - Challenges, Experiments and Perspectives}

\vspace{-15mm}	
\author{
	
	\IEEEauthorblockN{Chih-Hong Cheng, Frederik Diehl, Yassine Hamza, Gereon Hinz,  Georg N\"{u}hrenberg,\\ Markus Rickert, Harald Ruess, Michael Troung-Le}
	\IEEEauthorblockA{\textit{fortiss -  Landesforschungsinstitut des Freistaats Bayern}, Germany\\
		\texttt{\{lastname\}@fortiss.org}}
	
\vspace{-10mm}	
}

\maketitle

\begin{abstract}
	
We propose a methodology for designing dependable Artificial Neural Networks (ANN) by extending the concepts of understandability, correctness, and validity that are crucial ingredients in existing certification standards.
We apply the concept in a concrete case study in designing a high-way ANN-based motion predictor to guarantee safety  properties such as impossibility for the ego vehicle to suggest moving to the right lane if there exists another vehicle on its right.

\end{abstract}


\vspace{-2mm}
\section{Introduction}~\label{sec.introduction}
\vspace{-4mm}

The recent burst of applying artificial neural network (ANN) technologies has created an impact on applications such as autonomous driving. 
Although using ANN-based techniques had shown great promise (e.g., substantially superior image recognition~\cite{krizhevsky2012imagenet}) compared to classical  approaches, there have been huge barriers in using neural networks in safety critical domains  (e.g., report from NASA~\cite{bhattacharyya2015certification}). 

In this paper, we propose a methodology for enabling the usage of ANN by considering reasonable extensions for existing safety standards (Sec.~\ref{sec.certification.delta}).  We examine the technology readiness of our proposed methodology by applying a case study regarding highway motion prediction for autonomous driving (Sec.~\ref{sec.application.motion.prediction}), and address further research needs (Sec.~\ref{sec.conclusion.research.challenges}).

\begin{table*}[t]
	
	\centering
	\setlength\doublerulesep{0.1cm} 
	\caption{Extending the concept in certify safety-critical systems to new opportunities brought by neural networks.}
	\vspace{-3mm}
	\label{table:extension}
	\begin{tabular}{|p{2.6cm}|l|l|}
		\hline
		Implementation  	& Existing standard  & Fine-grained  specification-to-code traceability \\ \cline{2-3} 
		understandability	& Adaptation for ANN &  ($+$) Fine-grained neuron-to-feature traceability  \\ \hline\hline
		
		Implementation & Existing standard &  Verification based on testing and classical coverage criteria such as MC/DC \\ 
		
		\cline{2-3} 
		correctness	& Adaptation  for ANN  &   ($-$) coverage criteria such as MC/DC \\
		&      & ($+$) formal analysis against safety properties 
		\\ \hline\hline
		Specification & Existing standard &  Validation via prototyping, design-time analysis, 
		validity  and product acceptance test \\ 
		
		\cline{2-3} 
		validity & Adaptation  for ANN  &  ($+$) Validating data as a new type of specification  \\ \hline
	\end{tabular}
\end{table*}

\vspace{-2mm}

\section{Certification Considerations for Dependable Neural Networks}~\label{sec.certification.delta}

\vspace{-4mm}

For certification of safety critical systems, safety is established by rigorous~\emph{engineering processes} (i.e., these processes are defined in a way such that engineering complying to these processes can eliminate or prevent errors). Although it is more process-oriented than function-oriented, the basic  principle of (1) ensuring that the specification is correct and (2) ensuring that an implementation satisfies the specification is well perceived. 
Table~\ref{table:extension} summarizes three critical aspects over the underlying intention of certifying safety-critical systems, namely \emph{specification validity}, \emph{implementation understandability}, and \emph{implementation correctness}. 
\begin{itemize}
	\item The validity of specification is important to ensure that one ``builds the right system''. Several methods can be used in this regard, such as prototyping, design-time analysis and reviews, or product acceptance tests.
	\item The well-behaving of an implementation is captured by two aspects: (1) understandability via requirement-to-code traceability, and (2) correctness via extensive testing, with coverage criterion such as Modified Condition / Decision Coverage (MC/DC).
\end{itemize}

\noindent Although these approaches are valid for classical engineering using V-models, applying them on neural networks has created the following  issues:

\vspace{-2mm}

\begin{itemize}
		\item (\emph{Black-box structure}) For ANN-based systems, implementations consist of layers of neurons operating on and transforming high-dimensional vectors. This makes understandability arguments such as fine-grained requirement-to-code traceability difficult. 
	
	\item (\emph{Testing for correctness claims}) Depending on the activation function, applying traditional coverage-based approaches makes the system testing either trivially satisfiable or almost intractable.
	(i) When one uses $\tan^{-1}$ as the activation function, one only needs one test case to satisfy MC/DC  as there is no if-then-else branch in every neuron. 
	(ii) When one uses ReLU as the activation function, every neuron contains an if-then-else statement.  MC/DC is then intractable, as branching possibilities are exponential to the number of neurons.

	\item (\emph{Implicit specification}) For implementing systems using ANNs, the specification refers to a combination of data (which specifies input-output behaviors) as well as classical specifications for domain knowledge such as traffic or safety rules. The ``specification knowledge'' inside the data  is \emph{implicit}, compared to cases such as traffic rules.

\end{itemize}

\noindent Based on the above issues, Table~\ref{table:extension} further  summarizes our  considered additions towards safety certification of ANNs.

\begin{enumerate}[label=(\Alph*)]

\item (\emph{Neuron-to-feature understandability}) One should provide confidence regarding the meaning of a neural network by associating individual neurons with conditions (features) when it can be activated. 

\item (\emph{From testing to formal analysis}) The result of certification should provide (best effort) correctness claims over the (partially incomplete) classical specification, such as obeying traffic rules or ensuring road safety. As testing approaches its limitation, we suggest to apply formal methods such as static analysis or symbolic reasoning. 

\item (\emph{Validating the ``new specification''}) One needs to \emph{check the validity of the data}, to ensure that only sanitized data will be used in training. For examples such as autonomous driving, one needs to enhance raw data with sure guarantees such as no data containing risky driving has been introduced for training the maneuver of vehicles.

\end{enumerate}

\section{Case Study: Highway Motion Prediction for Autonomous Vehicles}~\label{sec.application.motion.prediction}

\vspace{-4mm}

\begin{figure}
	\centering
	\vspace{-4mm}
	\includegraphics[width=\columnwidth]{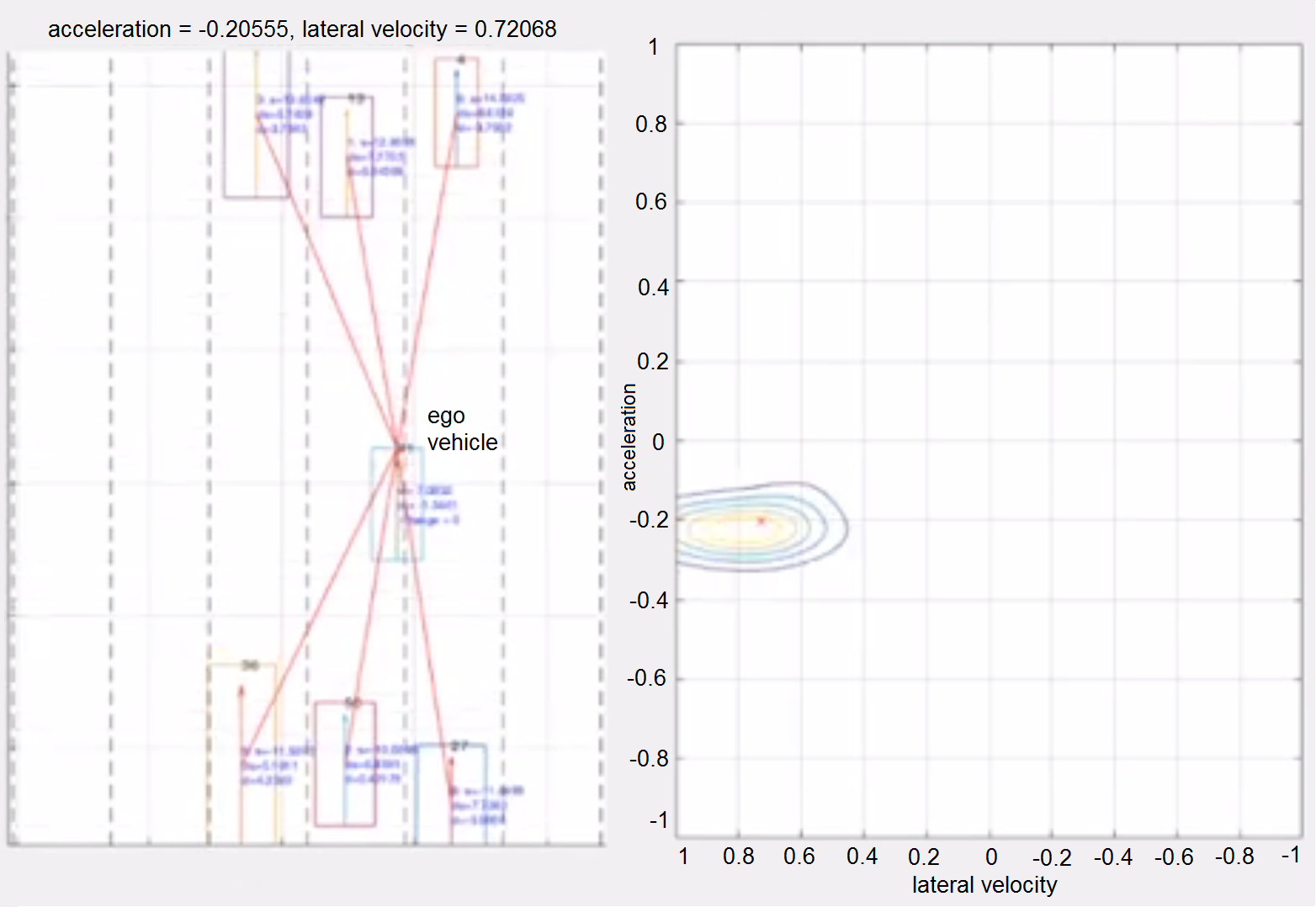}
	\vspace{-6mm}
	\caption{Simulation of the vehicle (left) and the switch-lane motion suggested by the neural network (right).}
	\label{fig:highway.simulator}
\vspace{-4mm}
\end{figure}

We outline how we applied the strategy above in verifying a highway overtaking ANN-based motion predictor used in autonomous driving  (developed by Lenz~et~al.~\cite{Lenz2017}).  Figure~\ref{fig:highway.simulator} provides a snapshot on the simulation of the vehicle. 

In Figure~\ref{fig:highway.simulator}, the ANN-based predictor takes three categories of inputs: (i) its own speed profile, (ii) parameters of its nearest surrounding vehicles for each orientation,  
and (iii) the road condition. The total number of input variables to the network  is~84. Given the current state of the perceived environment, it produces in real-time the probability distribution over all possible actions for a vehicle, characterized as a Gaussian mixture model.  The action of the ego vehicle is decomposed into two parts: (i) indicator over possible \emph{lateral velocity} (i.e., if it is feasible to switch lanes), and (ii) indicator over \emph{longitudinal acceleration} (i.e., if it is feasible to accelerate). In Figure~\ref{fig:highway.simulator}, the motion predictor on the right suggests to slightly decelerate  and to switch to left lanes, as the generated Gaussian mixture is within the lower left part.

\Comment{
\begin{figure}
	\centering
	\vspace{-3mm}
	\includegraphics[width=0.8\columnwidth]{fig/Cars_Feature}
	\vspace{-3mm}
	\caption{Tracking surrounding vehicles and road side conditions.}
	\label{fig:highway.sensor.map}
	\vspace{-3mm}
\end{figure}

}

One of the most critical safety requirements is to ensure that if there is a vehicle in the left of the ego vehicle, the predictor never suggests a large left velocity to the ego vehicle;  when such a scenario occurs, it may lead to crashes. In this example, it is regulated that the mean value of the probability distribution should be limited to certain threshold.  

Once we validated that the training data never contains such inputs (as in Sec.~\ref{sec.certification.delta} (A)), we perform formal verification (as in Sec.~\ref{sec.certification.delta} (B)) following the methodology developed by Cheng et al.~\cite{cheng2017maximum}, which encodes the structure of a neural network into a set of mixed integer linear constraints. With the technology we are able to successfully verify safety properties. Surprisingly, we have trained a couple of neural networks under the same data, but not all of them can guarantee the safety property (see Fig.~\ref{tab:opt} for a summary of verification results, being experimented on a Google VM with~12 Cores).

\begin{table}
		\vspace{-5mm}
	\centering
		\caption{Results of verifying ANN-based motion predictors.}
		\vspace{-3mm}
	\begin{tabular}{|l | p{4cm} | p{1.5cm} |}
		\hline
	ANN	& maximum lateral velocity, when exists a vehicle in the left & verification time \\ \hline
		$I_{4\times 10}$ & 0.688497 & 5.4s \\
		$I_{4\times 20}$ & 0.467385 & 549.1s \\
		$I_{4\times 25}$ & 2.10916 & 28.2s \\
		$I_{4\times 40}$ & 1.95859 & 645.9s \\
		$I_{4\times 50}$ & 1.72781 & 13351.2s \\
		$I_{4\times 60}$ & \texttt{n.a.} (unable to find maximum) & \texttt{time-out} \\ \hline	
				$I_{4\times 60}$ & Prove that the lateral
				velocity can  never be larger than $3
				\frac{m}{s}$ & 11059.8s \\ \hline		
	\end{tabular}

	\label{tab:opt}
\vspace{-5mm}
\end{table}

\vspace{-1mm}
\section{Concluding Remarks}~\label{sec.conclusion.research.challenges}

\vspace{-4mm}

The proposed certification methodology, during the case study, has also indicated
further research needs. 
\begin{enumerate}[label=(\roman*)]
	 \item During the study, we found that implementation understandability  can only be partially achieved  by technologies such as deconvolution~\cite{zeiler2011adaptive}.
		 \item Scalability of automated verification requires improvement (cf. Table~\ref{tab:opt} for required verification time). Recent results on quantized neural networks~\cite{hubara2016quantized} might make verification more scalable via an encoding to bitvector theories in SMT~\cite{de2008z3}. 
	 \item Apart from verification, another important direction is to consider training under known properties on the target function (known as  \emph{hints}~\cite{hints}), such as safety rules. 
\end{enumerate}

\end{document}